\documentstyle[psfig]{mn}
\input epsf 
\title{Hipparcos and the Age of the Galactic Disc}
\author[R. Jimenez, Chris Flynn \& Eira Kotoneva]
{Raul Jimenez$^{1}$, Chris Flynn$^{2}$ and Eira Kotoneva$^{2}$\\ 
$^1$Institute for Astronomy, University of Edinburgh, Royal Observatory 
Edinburgh, Blackford Hill, Edinburgh EH9 3HJ, UK\\ 
$^2$Tuorla Observatory, Piikki\"o, FIN-21500, Finland}

\begin{document}
\maketitle
\begin{abstract}

  We have used the Hipparcos satellite colour magnitude diagram to
determine the age of the Galactic disc.  We first measure the
metallicities of the clump stars using DDO photometry (H{\o}g \& Flynn
1997). We then use isochrones covering the range of disc metal
abundance and the morphology of the turn off and sub-giant region to
place constraints on the minimum age of the Galactic disc.  We derive
a minimum age of $11 \pm 2$ Gyr. In conjunction with the new ages
derived for globular clusters using the same method (Jimenez et al
1996) our results indicate that the delay between the formation of the
halo and the disc was up to 2-3 Gyr.  We show that a Reimers mass-loss
law is sufficient to explain the morphology of the red clump.

\end{abstract}

\section{Introduction}

  The age of the Galactic disc has been measured in the past using a
variety of independent methods.  The cooling rate of white dwarfs and
the lack of old, cool white dwarfs in the local disc places an upper
limit on the age of the Galactic disc of 6 to 10 Gyr, limited chiefly
by uncertainties in the cooling rates of white dwarfs (Liebert et al
1989, Winget et al 1987, Bergeron et al 1997).  The ages of evolved
F-stars can be determined from photometry and isochrones, indicating
that the disc is between 10 and 12 Gyr old (Edvardsson et. al. 1993).
Radioactive dating (e.g. isotope ratios) establishes a method for
measuring the age of the disc although technically quite difficult but
with a lower limit of approximately 9 Gyr for the age of the Galactic
disc (Butcher 1987, Morell et al 1992).  A fourth method use the lower
locus of the red giant branch in the colour magnitude diagram: this
method places a firm lower limit of $8\pm1$ Gyr on the age of the disc
through a comparison of the colour magnitude diagram of field G and K
giants with the giant branch of the old open cluster NGC 188 (Janes
1975, Wilson 1976, Twarog and Anthony-Twarog 1989).

  In this paper we measure the age of the Galactic disc from the field
stars, by comparing the colour magnitude diagram of the local disc as
measured by the Hipparcos mission with a range of stellar isochrones.
Constraints on the isochrones are obtained through the use of accurate
metallicities of the G and K giants from H{\o}g \& Flynn (1997) and
for G and K dwarfs from Flynn and Morell (1997).  Our lower limit to
the disc age comes essentially to the fit of the isochrones in the
sub-giant region to the Hipparcos data. Having constrained the age in
this way, we then study how mass-loss determines the red clump
morphology and its relation with the red giant branch morphology.  Our
method is based on the one developed by Jimenez et al (1996) to study
the ages of globular clusters.

\begin{figure*}
\centerline{
\psfig{figure=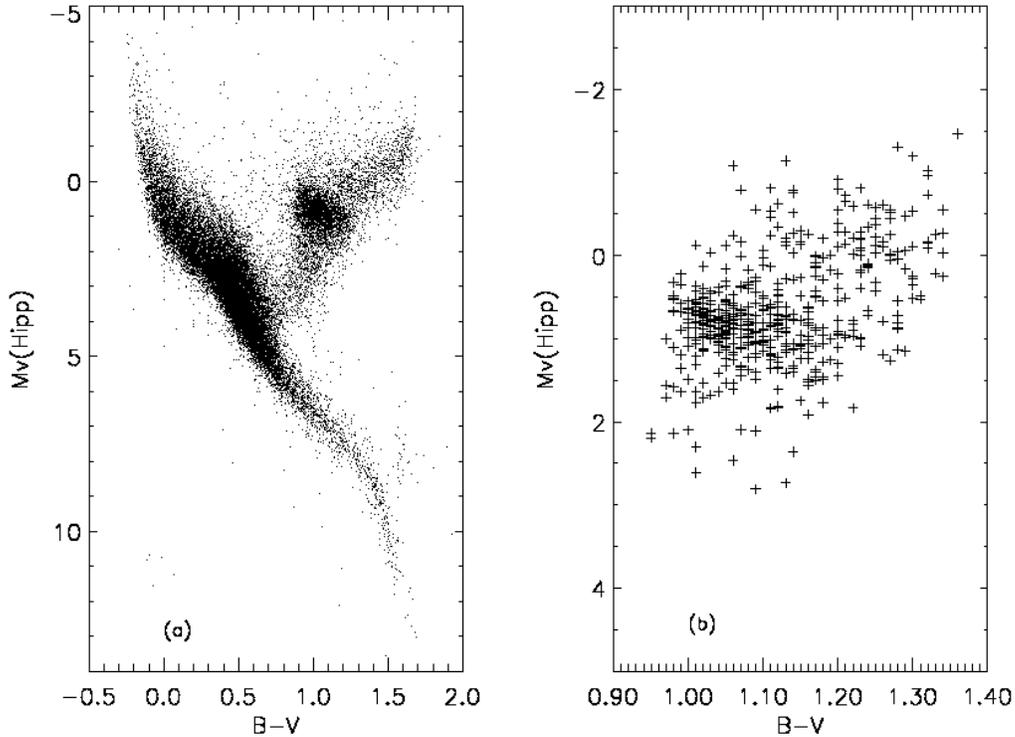,height=10cm,angle=90}}
\caption[]{Colour-magnitude diagrams for the Hipparcos catalogue with stars 
whose parallaxes are with errors $<0.15$ (left panel) and the same for the 
sample of giants for which metallicities have been measured by H{\o}g \& Flynn 
(1997).}
\end{figure*}

  This paper is organised as follows: In section 2 we describe the
data from the Hipparcos satellite. In section 3 we describe the models
we use to estimate the age of the disc and the morphology of red
clump.  We discuss our derived age in section 4 and conclude in
section 5.

\section{COLOUR MAGNITUDE DIAGRAM AND METALLICITIES}

 Our study is based on the colour magnitude diagram (CMD) of local
stars observed by the {\it European Space Agency's} Hipparcos
Satellite.

 Data from Hipparcos were released in July 1997.  We prepared a CMD
for all stars for which the parallax, $\pi$ had been measured to
better than 15\%.\footnote{We note that the typical error on the
parallaxes for these giants is only 8\% (as discussed in H{\o}g and
Flynn 1997) or an absolute magnitude error of 0.15 mag. This is so
accurate that no Lutz-Kelker type corrections to the parallaxes were
necessary (see H{\o}g and Flynn 1997)} The resulting subsample in the
entire Hipparcos catalogue is plotted in Fig. 1(a). The rapidly rising
red giant branch (RGB) is populated by first ascent giants as well as
the remarkably clear clump (or He core burning) giants at an absolute
magnitude of $M_V\approx 0.8$.

  Stellar evolutionary state of the giant branch is quite sensitive to
metal abundance, and we require estimates of [Fe/H] for the giants to
determine the disc age.  H{\o}g \& Flynn (1997) have analysed the
metal rich K giants in Hipparcos with ($0.95 < B-V < 1.4$) in order to
calibrate a photometric indicator of K giant absolute magnitude. The
photometric indicator uses intermediate band DDO photometry (Janes
1975), with which metal abundances for the individual giants in our
sample have been derived using the Janes (1975, 1979) method.  The CMD
of these stars is shown in Fig 1(b). We note that the H{\o}g \& Flynn
sample was limited to metal rich giants [Fe/H] $ > -0.5$ and $ 0.95 <
B-V < 1.35$, meaning that the bluer, metal poor clump stars are not
included.  Stars with [Fe/H] $< -0.5 $ are either from the thick disc
or halo, with different kinematics and possibly formation history to
the disc, which is why we exclude them from the sample.  Finally,
there are few intrinsically luminous red giants in the sample, because
we have limited ourselves to stars with parallax measurements of
better than 15\%. This tends to exclude intrinsically bright ($M_V <
-1)$ giants. This causes us no concern either because the clump stars
are well below this limit. 

\section{Age measurement}

  Nearby K giants provide a snapshot of the chemical history of the
local Galactic disc. Over the life-time of the disc, successive
generations of main sequence stars have formed, and we see a
particular selection of them today by age, mass and metal abundance
passing through the giant phase. The measurement of age from the
Hipparcos CMD is therefore somewhat more involved than in open or
globular clusters, where one has a homogeneous population of stars.

\begin{figure*}
\centerline{
\psfig{figure=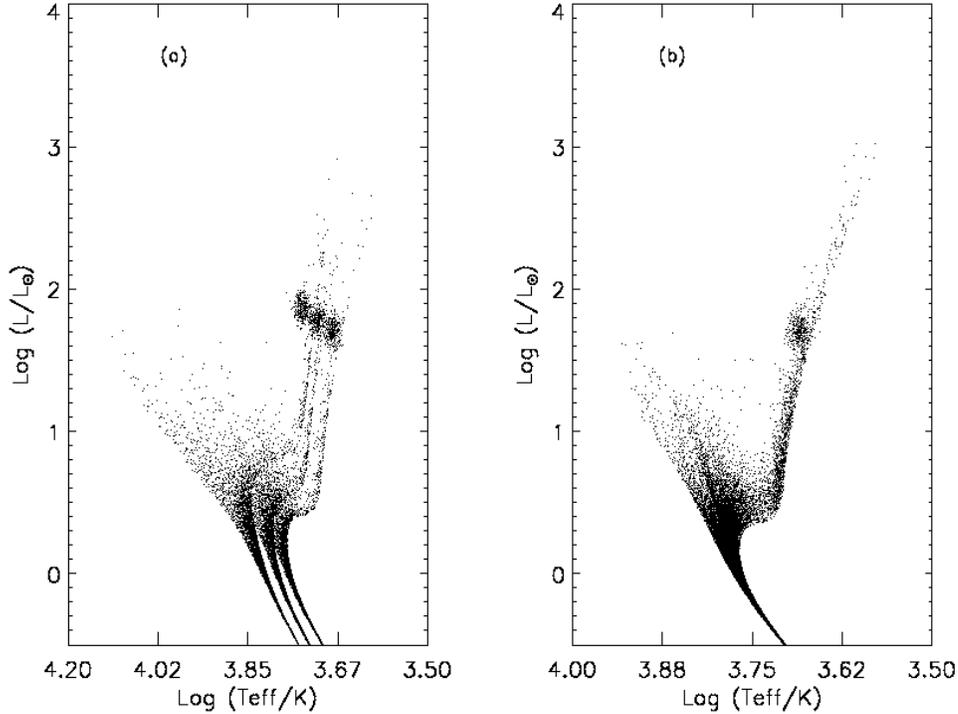,height=10cm,angle=90}}
\caption[]{Synthetic CMD for a different star formation histories and
metallicities. The left panel shows the synthetic CMD for a population
with 3 metallicities ([Fe/H]$=0.0$, [Fe/H]$=-0.7$ and [Fe/H]$=-2.0$),
for each of these the star formation history with an initial burst
that decayed exponentially with $\tau= 2$ Gyr, the total age of the
population is 10 Gyr.  The right panel shows a population with a
constant star formation rate over 10 Gyr. It transpires from both
panels that a spread in the RGB and the main sequence can only be due
to the spread in metallicity in the disc.}
\end{figure*}

  One of the most remarkable features of the Hipparcos CMD is the
Horizontal Branch clump (see Fig. 1). Contrary to the situation in
globular clusters where the HB is a sharply defined {\it horizontal
and thin} line, the equivalent in the field is much thicker in
absolute magnitude.  From Figure 1 one can see that the clump is about
0.7 magnitude broad (FWHM), much more than the scatter in the absolute
magnitudes due to measurement error which is only 0.15 mag.

  The vertical and horizontal extent of the clump and the width of the
giant branch seen in the Hipparcos CMD are caused by the range of
stellar age and metallicity currently passing through this
evolutionary phase. We illustrate this in Figure 2, where we plot
synthetic CMDs for different scenarios of star formation and
metallicities of a population representative of the local disc. We
have computed these diagrams using the most recent version of our
synthetic stellar population code (Jimenez et al. 1997). A detail
description of the code has been given in Jimenez et al. 1997 to which
we refer the reader for full details.  Fig.~2(a) shows synthetic
stellar populations in which star formation took place in a initial
burst of $1 \times 10^6$ years and then decayed exponentially with
$\tau=3$ Gyr. It was calculated for three different metallicities:
[Fe/H] $=0.3$, [Fe/H]$=0.0$ and [Fe/H]$=-0.7$ (from right to left). In
all cases $dY/dZ=2.5$. Fig.~2(b) shows a model where metallicity was
kept fixed to [Fe/H]$=0.0$ and we used continuous star formation to
model the disc population. In both cases the age of the population was
chosen to be 10 Gyr. Although both of these model types have plausible
star formation laws, no model at a single metallicity would provide a
good match to the Hipparcos data.  Not unexpectedly, the metallicity
spread is an important factor in the CMD, causing scatter along the
main sequence and giant branch{\footnote{It is possible that
star-to-star variations in the mixing length parameter $\alpha$ are
causing some of the scatter. However, this would be in contradiction
to a recent analyses of variations of $\alpha$ in globular clusters
(c.f. Jimenez et al. 1996), and we do not consider this a likely
scenario amongst disc stars}}.

  As is well known, metallicity has a strong effect on the colour of
clump stars formed. We show in Fig. 3 a calculation using the Jimenez
et al code of the zero age horizontal branch for different masses at
several metallicities. It is worth noticing three features:

\begin{enumerate}

\item For masses larger than 0.8 $M_{\odot}$ (i.e. ages less than
about 16 Gyr) the HB is {\it not horizontal} but {\it vertical}. This
means that a well defined red limit exists to the clump for masses
between 0.8 and 1.3 $M_{\odot}$ (i.e. ages between 16 and 2 Gyr).

\item The higher the metallicity the redder the red limit. Therefore
the metallicity distribution of a stellar population can be estimated
from the colour distribution of the clump stars.

\item The lower locus of the sub-giant branch is occupied by the
oldest stars in the population and also the most metal rich.  A
minimum age estimate for the Galactic disc can be obtained from these
stars if accurate metallicities and distances are known.

\end{enumerate}

\begin{figure*}
\centerline{
\psfig{figure=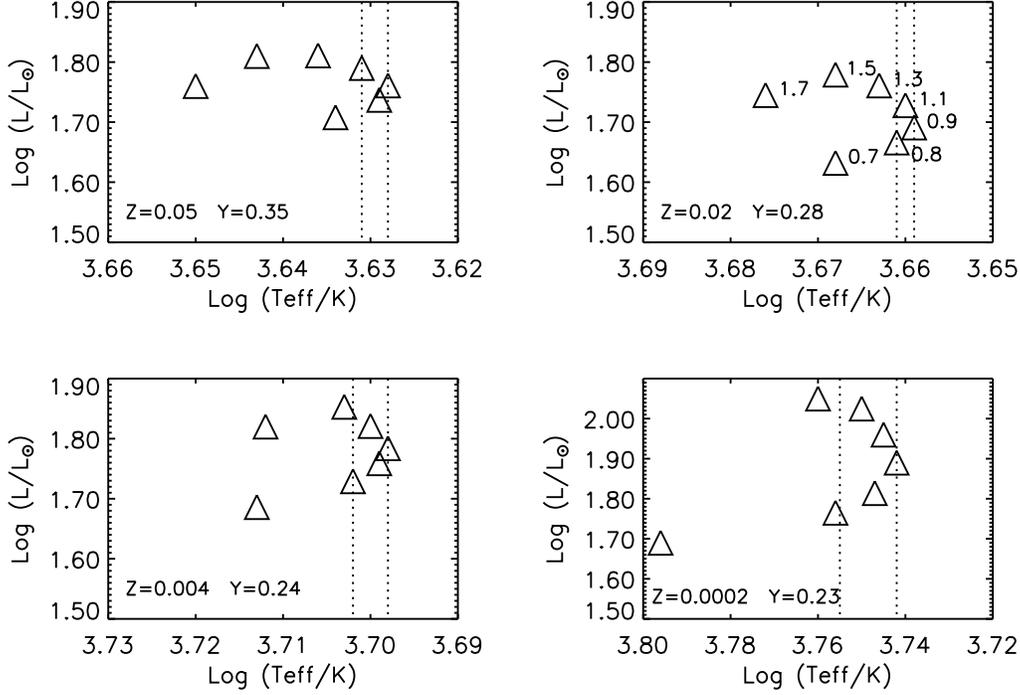,height=10cm}}
\caption[]{The position of the ZAHB for four different metallicities
and different masses. There is a well defined red limit that depends
on the metallicity of the stars. For a range of masses between 0.8 and
1.3 $M_{\odot}$ (i.e ages between 16 and 2 Gyr), the HB is {\it
vertical}.}
\end{figure*}

  In Fig. 4 (left panel) we have superimposed our calculated red
limits of the clump for the four metallicities analyzed in Fig. 3
(left to right: [Fe/H]$=-2.0, -0.7, 0.0, 0.3$) on the Hipparcos
data. The metallicities of the stars are seen to be in the range $-0.6
< $[Fe/H]$ < 0.2$, in good accord with the known abundance
distribution of the disc (see e.g. Freeman 1987). Note that the lines
show the red limit of the clump rather than the mean colour at a
particular metallicity. 

  In the case of these bright disc giants, detailed individual
abundances are available from H{\o}g and Flynn (1997), and we make use
of this information in what follows.  In order to determine the
metallicity content of the clump, we isolate it using the two dashed
lines shown in figure 4 and plot the metallicities of the stars as a
function of colour (Fig.~5). Most of the stars in this cut are true
clump stars, although some normal ascent giants are also moving
through this region. Clump stars dominate however, as they can be
still clearly seen in Fig.~5 despite the background signal from normal
first ascent giants.  There is a trend along the clump of increasing
metallicity with increasing colour as expected from the results shown
in Figure 3.

\subsection{Disc age from the reddest clump stars and the morphology 
of the sub-giant branch}

  We now estimate the disc age as follows. Consider giants of solar
metallicity [Fe/H] $=0.0$, in Fig 5.  At this metallicity, the red
limit of the clump appears at $B-V \approx 1.17\pm0.02$. The mean
absolute magnitude of such giants (from Hipparcos) is $M_V = 0.7$.
This defines a fiducial point for solar metallicity isochrones.  We
can construct two further fiducial points, for giants with
[Fe/H]=$-$0.5 and [Fe/H]=0.3, i.e. the lowest and highest disc
metallicities (see also Fig.5).  The fiducial points are shown as
crosses in Figure 6.

  We now construct isochrones of different ages (8,11,13 and 15 Gyr)
for the above metallicities, and overlay them on the Hipparcos CM,
shown in Figure 6 and 7. The isochrones all pass very close to the fiducial
points, giving us some confidence that the the metallicity scale and
colour transformations of the isochrones is good.  As expected, the
feature that is age sensitive is the luminosity of the main sequence
turn off and the morphology of the sub-giant branch (see Fig.~6).

\begin{figure}
\centerline{
\psfig{figure=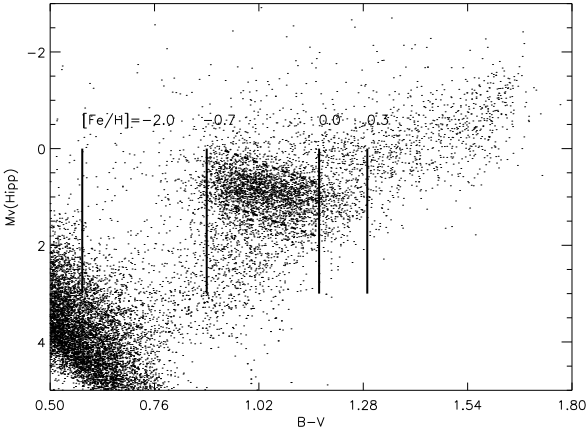,height=6cm,angle=90}}
\caption[]{The CMD for the Hipparcos catalogue and the H{\o}g \& Flynn sample. 
Superimposed are the red limits defined by the theoretical ZAHB models
for different metallicities (form right to left: $Z=2Z_{\odot}$,
$Z=Z_{\odot}$, $Z=Z_{\odot}/5$ and $Z=Z_{\odot}$/100). It is clear
that the metallicity of the HB clump is well constrained between solar
and 1/5 solar metallicity.  Dashed lines are used to isolate clump
stars plotted in figure 5 (as discussed in section 3).}
\end{figure}

  We now set limits on the age of the disc. First, one sees from
Fig.~6 that ages younger than 8 Gyr can be ruled out. If we examine
the [Fe/H] $= 0.3$ isochrones in Fig 6(a), then the 8 Gyr isochrone
forms a reasonable lower locus to the faintest Hipparcos
sub-giants. This sets the minimum disc age to be 8 Gyr, since a
younger age could only be obtained for isochrones more metal rich than
[Fe/H] $= 0.3$, but this is the practical maximum disc metallicity
(see Figs. 4 and 5). We can show that the disc must in fact be older
than 8 Gyr by examining the isochrone set for solar abundance (Fig
6b).  A maximum disc age of 8 Gyr seems a poor fit for solar
metallicity stars. If the oldest disc stars were only 8 Gyr old, then
all the stars below the 8 Gyr solar metallicity isochrone would have
to be explained as stars with [Fe/H] $> 0.0$.  This is in
contradiction with two facts: from Fig.~5 (and also from Fig.~4) it is
clear that only about 10\% of the stars in the red clump have
metallicities larger than solar, and only a fraction of these will be
as old as the disc. Even at a metallicity of [Fe/H]=0.3 (Fig.~6 panel
(a)) there are sub-giants stars below the 8 Gyr isochrone that are
only fitted at an age of 10-11 Gyr.  A consistent picture is obtained
by adopting 11 Gyr as the minimum age of the disc. In this case, the
solar abundance isochrones fit the sub-giant region of the Hipparcos
data well and there is no discrepency with the super-solar
isochrones. Of course, disc stars can be comfortably older than 11 Gyr
as long as they are of less than solar metallicity, as seen in Figure
6(c). We conclude that the minimum disc age is 11 Gyr.



\begin{figure}
\centerline{
\psfig{figure=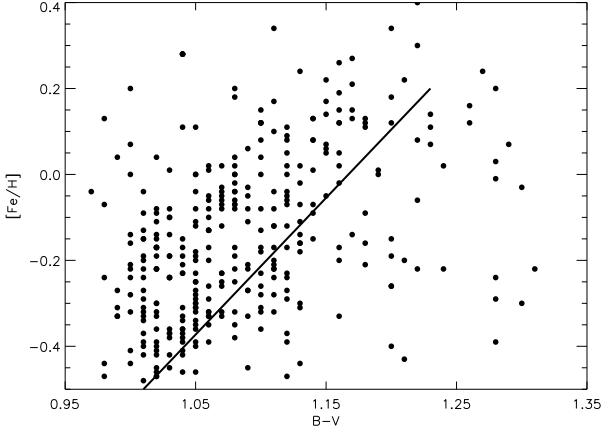,height=6cm,angle=90}}
\caption[]{Plot of colour versus metallicity for the clump stars as
isolated by the dashed lines in Figure 4. This plot shows mostly clump
stars, although some first ascent giants which happen to be passing by
the clump also appear in the plot. There is a correlation between the
colour and metallicity of the clump stars, as expected from the models
and as illustrated in Figure 3. The solid line marks the approximate
position of the red edge of this trend, i.e. the colour of the reddest
clump stars as a function of metallicity.}
\end{figure}
 
\subsection{Error estimate of the disc age}

  We now discuss in turn the main sources of error in the age
estimate, which are the metallicity scale and the colour
transformations of the isochrones.

  Firstly, a potential source of systematic error in our age estimate
would be a difference between the metallicity scales of the isochrones
and the clump giants (which are used to establish the fiducial points
in Figure 6). The isochrones are in fact good fits to these fiducial
points, indicating that the metallicity scales are not at odds with
each other.  However, the metallicities of the clump stars come from
DDO photometry (Janes 1975) which is calibrated by spectroscopic
analysis of field giants.  Field K giants show a change in kinematics
from disc to thick disc at an abundance of [Fe/H] $=-0.5$, whereas the
same kinematic change appears in F and G dwarfs in the Edvardsson
et. al. (1993) sample of disc stars to be closer to [Fe/H] $=
-0.4$. We therefore consider the possibility that the abundance scale
has a systematic error of order 0.1 dex. From Figure 5, a systematic
shift in the metallicity scale of 0.1 dex is equivalent to a colour
shift of 0.05 mag in the position of the fiducial points. Under these
circumstances the isochrones would no longer fit the fiducial points,
but if we choose to ignore these points, then a colour shift of 0.05
mag results in a change in the age estimate of order 1 Gyr. This can
be seen by examining the colour differences between the isochrones in
the sub-giant region of Fig 7.

  Secondly, the colours of our isochrones are dependent upon having
the right colour transformations. We have some confidence in the
isochrones that pass through the clump region, since they are a good
fit to the observed red limit of the clump. Despite fitting in the
clump region, the isochrone colours could be systematically offset in
the sub-giant region or main sequence.  We have investigated this as
follows. Metallicities for G and K dwarfs in the Gliese catalog have
been measured by Flynn and Morell (1997). All of these stars now have
parallaxes measured by Hipparcos.  We show in Figure 7 the Flynn and
Morell G and K dwarfs with abundances in the range $-0.6 < $ [Fe/H] $
< -0.4$ as squares, and we show stars with $-0.05 < $ [Fe/H] $ < 0.05$
as crosses. The three sets of isochrone from Fig. 6 are also
displayed.  Relative to the solar abundance dwarfs, our solar
metallicity isochrone is a little too red, but by not more than 0.1
mag.  Our [Fe/H] $= -0.5$ isochrone matches the data better, although
it is a little too blue for the brighter dwarfs. A small number of
dwarfs in the range $0.2 < $ [Fe/H] $ < 0.4$ were also examined but
are not shown for clarity. From these our [Fe/H] $ = 0.3$ isochrone
appears to be too red by 0.1 mag.  We conclude that a systematic
colour error is possible in the isochrones in the sub-giant and dwarf
region of up to 0.1 mag.  Examining the isochrones in Fig 6, one can
see that in the sub-giant region, shifting the isochrones by 0.1 mag
to the red is equivalent to increasing the disc age estimate by 2 Gyr.

\begin{figure*}
\centerline{
\psfig{figure=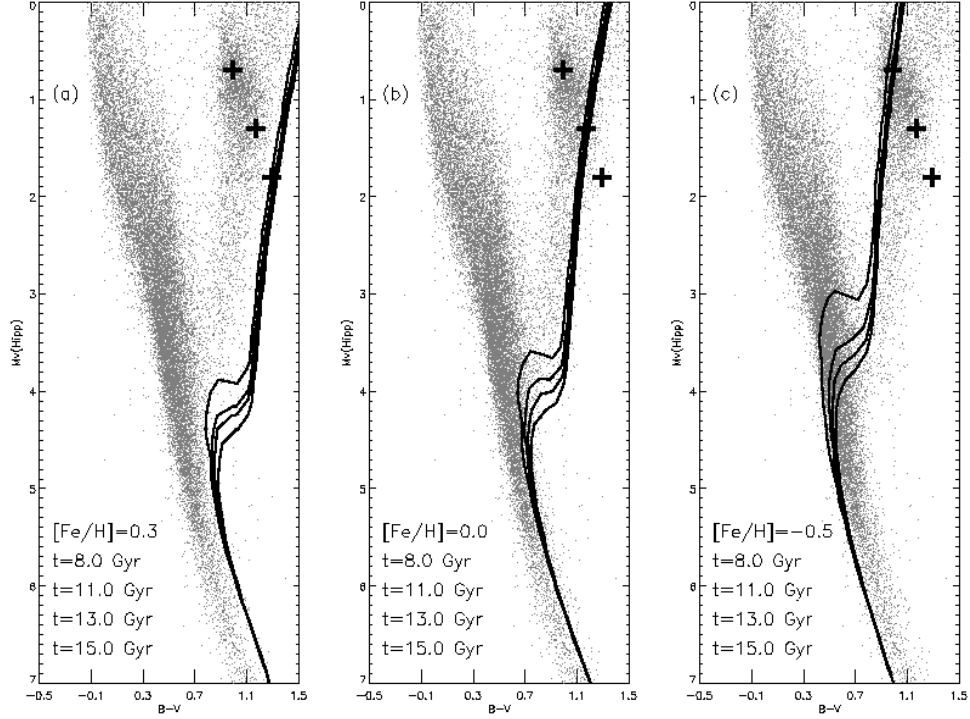,height=10cm,angle=90}} 
\caption[]{Colour magnitude diagrams of the Hipparcos data, as in
Figure 1, overlayed by isochrone fits. The crosses show the colours of
the reddest stars in the clump as a function of metallicity (see
Figure 5), for three metallicities [Fe/H] $=0.3, 0.0$ and [Fe/H] =
$-0.5$, chosen to bracket the disc abundance distribution. A minimum
disc age of $11\pm2$ Gyr is derived from these plots and is discussed
in detail in section 3}
\end{figure*}

  An obvious concern in our method is that we have assumed that the
helium enrichment follows $dY/dZ=2.5$. If we adopt other values for
$dY/dZ$ the absolute values of the age of the Galactic disc will
change accordingly, but so will the age of the Sun.  For example
changing $dY/dZ$ at solar metallicity to 1.1 would imply a change in
the disc age from 10 Gyr to 14 Gyr, but this would change the Sun's
age to 6 Gyr. Our age scale for the disc is thus firmly tied to a
solar age of 4.5 Gyr, and the ages of the disc and globular clusters
(Jimenez et al 1996) are all relative to this.

  We conclude from the discussion of metallicity and colour effects in
this section that our age error is of order 2 Gyr.

\subsection{The morphology of the Red Clump}

\begin{figure}
\centerline{
\psfig{figure=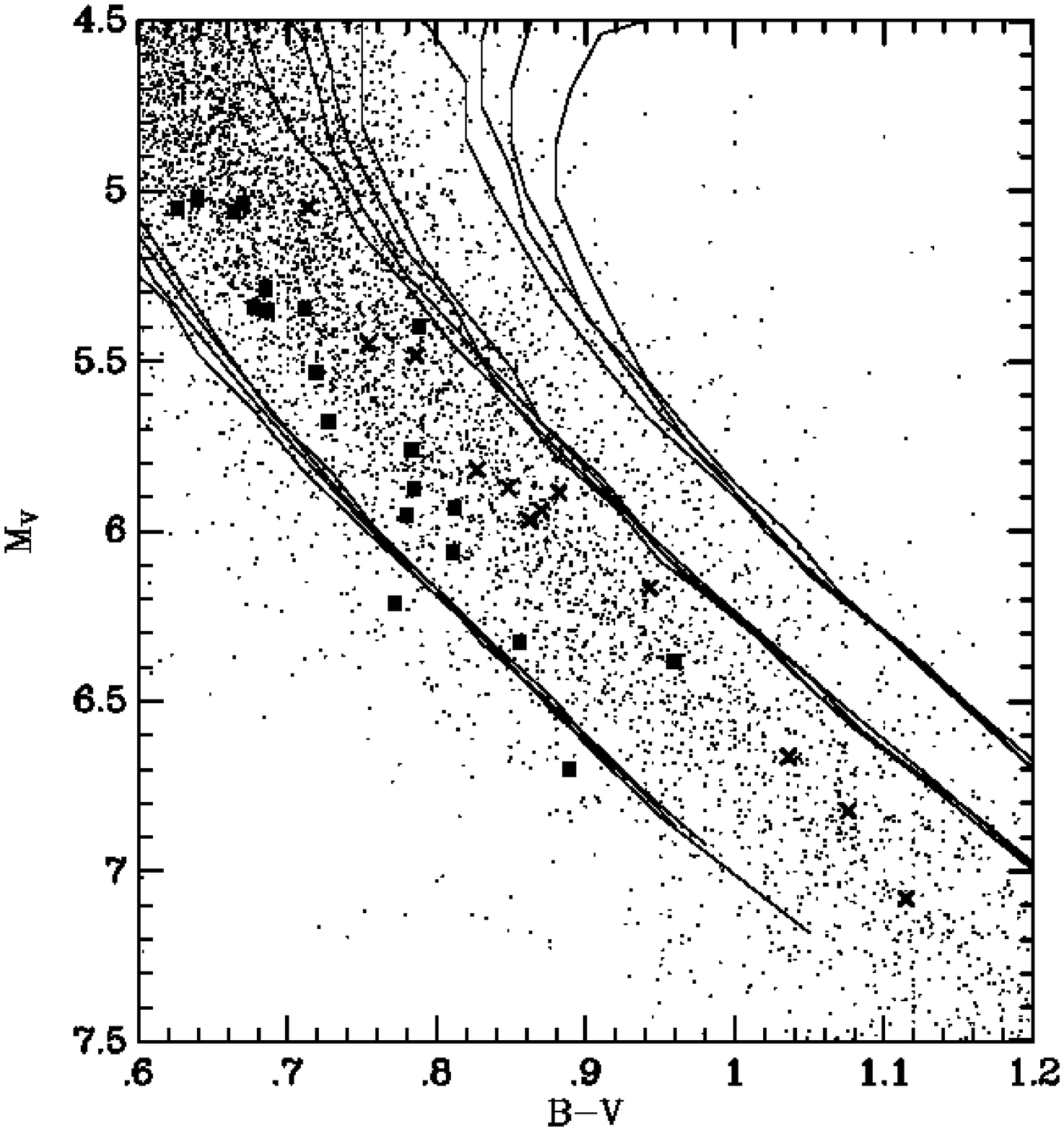,height=10cm,angle=0}}
\caption[]{Colour magnitude diagram of the Hipparcos data, showing the
lower main sequence, overlaid by isochrone fits. G and K stars on the
main sequence for which abundances are available from Flynn and Morell
(1997) have been plotted in two metallicity ranges $-0.6 < $ [Fe/H] $
< -0.4$ (squares) and $-0.05 < $[Fe/H]$ < 0.05$ (crosses). The lower
set of isochrones have [Fe/H] $= -0.5$ and match the data quite well.
The middle set of isochrones has [Fe/H] $ = 0.0$ and appears too red
by about 0.07 mag. The upper set of isochrones has [Fe/H] $ = 0.3$,
and it too is too red by about 0.1 mag (the comparison stars have not
been shown for clarity). The uncertainty in the position of the
isochrones of 0.1 mag sets a limit of about 2 Gyr to the accuracy with
which we can measure the disc age.}
\end{figure}

  The age estimate of $11 \pm 2$ Gyr above is partially based on the
reddest stars in the clump as a function of metallicity.  The
thickness of the clump in absolute magnitude could itself be used to
measure the age, if it were not for the uncertain amount of mass-loss
taking place.  Specifically, if we neglected mass-loss, the luminosity
of the least luminous stars of the red clump of Fig 3 would led to a
minimum age estimate of the disc of circa 30 Gyr, much older than our
age estimates for the globular clusters of about 13 Gyr {\it using the
same method} (Jimenez et al 1996).  We conclude that mass loss plays
an important role in field red clump's morphology, as it does in the
HB in globular clusters (Jimenez et al. 1996). The only alternative
explanation would be cumbersome variations of the Helium content in
the Galactic disc that require fine tuning.

\begin{table*}
\begin{tabular}{cccccccc}
\hline
\hline
RGB Mass/M$_\odot$ ($\eta=0.0$ [Fe/H]=0.0  Y=0.28) & 0.7  & 0.8 & 0.9 & 1.0 & 1.1 & 1.2 & 1.5 \\
Age/Gyr                     & 42.7 & 26.1& 16.9& 11.5& 8.0 & 5.8 & 2.6 \\ 
Clump Mass/M$_\odot$                  & 0.7  & 0.8 & 0.9 & 1.0 & 1.1 & 1.2 & 1.5 \\
\hline
RGB Mass/M$_\odot$ ($\eta=0.4$ [Fe/H]=0.0  Y=0.28) & 0.7    & 0.8 & 0.9 & 1.0 & 1.1 & 1.2 & 1.5 \\
Clump Mass/M$_\odot$                  & Manque & 0.61& 0.75& 0.88& 1.0 & 1.12& 1.43 \\
\hline
RGB Mass/M$_\odot$ ($\eta=0.8$ [Fe/H]=0.0  Y=0.28) & 0.7      & 0.8 & 0.9 & 1.0 & 1.1 & 1.2 & 1.5 \\
Clump Mass/M$_\odot$                  &  Manque  & Manque & 0.52 & 0.72 & 0.87 & 1.0 & 1.3 \\ 
\hline \hline
RGB Mass/M$_\odot$ ($\eta=0.0$ [Fe/H]=$-$0.5  Y=0.24) & 0.7  & 0.8 & 0.9 & 1.0 & 1.1 & 1.2 & 1.5 \\
Age/Gyr                        & 33.9 & 20.7 & 13.4 & 9.1 & 6.4 & 4.6 & 2.0 \\
Clump Mass/M$_\odot$                     & 0.7  & 0.8 & 0.9 & 1.0 & 1.1 & 1.2 & 1.5 \\
\hline
RGB Mass/M$_\odot$ ($\eta=0.4$ [Fe/H]=$-$0.5  Y=0.24) & 0.7  & 0.8 & 0.9 & 1.0 & 1.1 & 1.2 & 1.5 \\
Clump Mass/M$_\odot$                  & 0.51 & 0.65 & 0.78 & 0.90 & 1.01 & 1.10 & 1.41 \\
\hline
RGB Mass/M$_\odot$ ($\eta=0.8$ [Fe/H]=$-$0.5  Y=0.24) & 0.7    & 0.8 & 0.9 & 1.0 & 1.1 & 1.2 & 1.5 \\
Clump Mass/M$_\odot$                  & Manque & Manque & 0.62 & 0.77 & 0.91 & 1.04 & 1.3 \\
\hline \hline
\end{tabular}
\caption[]{Derived disc ages for different clump masses, mass-loss
rates, metallicities and He abundance $Y$. Manque means that the stars
evolve directly from the RGB into a white dwarf.}
\end{table*}

  We can fix the age of the disc as derived in the previous section at
$11\pm2$ Gyr, and determine the amount of mass loss in the clump
instead.  We have isolated, using the H{\o}g \& Flynn (1997)
abundances, the red end of the clump for two particular metallicities
[Fe/H]$=0.0$ and $-0.45$ (see Fig. 8). A comparison with Fig. 3 shows
that the red clump covers a mass range between 0.8 and 1.5 M$_{\odot}$
in both cases.  Fixing the age at 11 Gyr, the mass of the stars on the
RGB should be about 1 M$_{\odot}$. Since the lowest luminosity stars
in the red clump have masses of about 0.8, they must have lost 20\% of
their mass. This is in excellent agreement with what is expected from
the (empirical) Reimers mass-loss law (Reimers 1975). In table 1 we
show the predictions for a Reimers mass law with the $\eta$ parameter
(see Reimers 1975) in the range 0.0 to 0.8.

  Using the number of stars in the clump for each metallicity it is
possible to predict the distribution in the $\eta$ parameter. From the
bottom panel of Fig. 6 we conclude that $\eta$ is peaked at about 0.4,
which is in excellent agreement with measurements in the field and the
morphology of the HB in globular clusters (Jimenez et al. 1996). We
remark that the underlying population of first ascent giants in the
red clump is not of concern for this analysis since it only adds an
offset to the bottom diagrams of Fig. 6 but does not change
significantly the shape of the distribution.

  We note that the morphology of the red clump cannot be due to
star-to-star variations in the mixing length parameter. This is
because the red clump mass distribution is vertical for giants of the
typical age and mass found in the disc and is therefore unaffected by
variations of this type.

\begin{figure*}
\centerline{
\psfig{figure=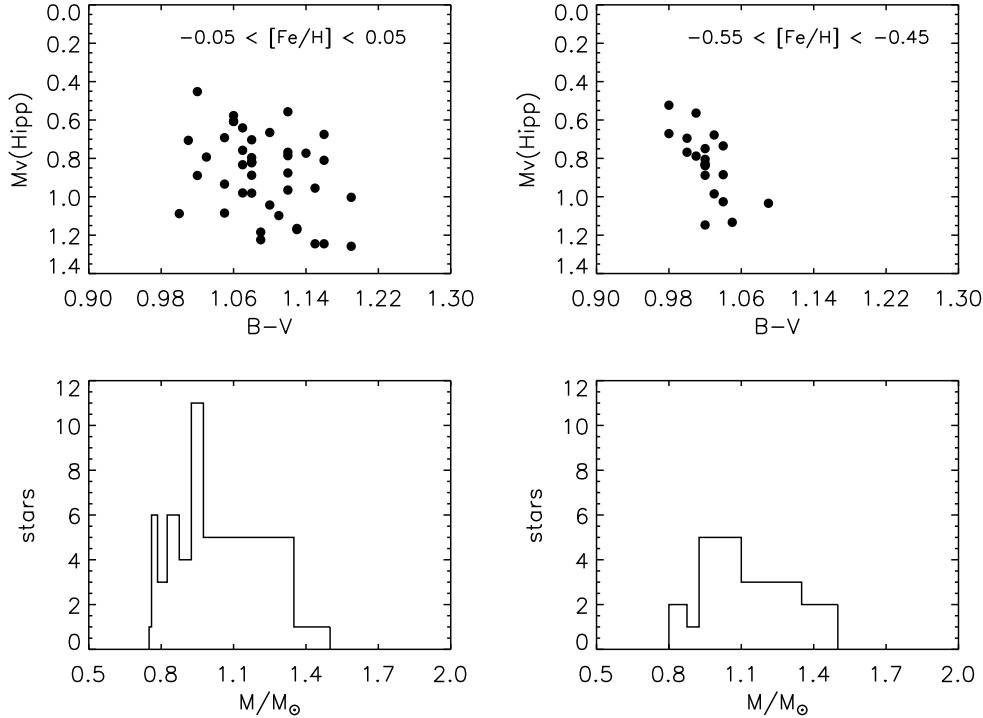,height=10cm,angle=90}}
\caption[]{The Upper panels show colour versus absolute magnitude for
stars in the clump for two metallicity ranges. For each star a mass
can be estimated using the models plotted in Figure 3. The mass
distributions are in the lower panels (note the use of non-equal bin
sizes for practical reasons). These distributions of mass in the red
clump can be well modelled by the conventional Reimers mass-loss law}
\end{figure*}
\section{Discussion}

  Our result is that the disc has a lower age limit of $11\pm2$ Gyr.
We can compare this estimate to that obtained for the field F and G
dwarfs for which age estimates can be made via distances and isochrone
fitting (Edvardsson et al 1993). The oldest disc stars ([Fe/H] $ >
-0.5$) in their sample are as old as 12 Gyr. Recently, the ages of
their stars have been redone using the Hipparcos parallaxes (Ng and
Bertelli 1997) and these authors confirm this age for the oldest disc
stars. All stars which they find are older than 12 Gyr have [Fe/H]
$<-0.5$ and are kinematically members of the thick disc (Edvardsson et
al 1995, Figure 16). Hence this age is consistent with our
determination of $11 \pm 2$ for the disc ([Fe/H] $> -0.5$) stars.

  We now compare this to the ages derived for the globular clusters in
the Jimenez et al (1996) sample.  There are six halo globular clusters
(we define this as [Fe/H]$ \le -1.2$: M3, M5, M22, M68, M72 and M92)
and two disc globular clusters (M107, 47 Tuc) in the sample. The mean
age of the six halo clusters is 13.0 Gyr, with a remarkably small
dispersion of 0.3 Gyr; if these clusters are at all representative
then the halo formed rapidly at this time, 13 Gyr ago. This is a
relatively young age for the halo, but recently substantial support
for this age scale has come from the analysis of Hipparcos parallaxes
for nearby sub-dwarfs (Reid 1997), which shows that the distance
modulii to globular clusters have been traditionally underestimated
and the ages overestimated. Reid estimates that the average age of the
globular clusters is less than 14 Gyr, and may be as low as 12 Gyr.

  Our results do not favour a substantial age spread in the globular
clusters (Bolte 1989), which has long been a proposal for explaining
the large range of horizontal branch colour seen in globular clusters,
the so-called second parameter problem. Rather, the remarkably co-eval
formation times of these clusters is in good agreement with the idea
that the globular cluster mass-scale is an important one and may
represent the dominant mode by which stars formed at that epoch
(Padoan, Jimenez and Jones 1997).

  The two disc clusters have ages of 12.0 and 11.5 Gyr, while the disc
age as measured in this paper is $11\pm 2$ Gyr. We should be careful
to draw the distinction between the disc age measured in this paper
near the sun (at a Galactocentric distance of 8 kpc) and the ages of
disc globular clusters, which are mostly closer to the center of the
Galaxy than 4 kpc, where the disc may be older than at the solar
circle. In any case, the sequence of ages in this small sample
indicates that there was a delay of 2-3 Gyr between the formation of
what now traces the halo and the time when most of the gaseous phase
of the proto-galaxy had settled into place in the disc.  This is very
much in accord with a relatively gentle bottom-up picture of the
formation of the Galaxy (cf Padoan, Jimenez and Jones 1997, Reid
1997), as well as the results of deep imaging with the Space Telescope
(Pascarelle et al 1996).

  Discs are relatively easy to disrupt via accretion (Toth and
Ostriker 1992). At an age of 11 Gyr, it is unlikely that the disc has
suffered a major accretion event during this time.  The formation
redshifts of the Galactic disc in two representative cosmologies are
$z=1.8$ in a closed Universe (with $H_0=48$ km s$^{-1}$ Mpc$^{-1}$)
and $z=3.2$ in an open Universe ($\Omega=0.3$, $H_0=65$ km s$^{-1}$
Mpc$^{-1}$). Rotation curves and morphologies of disc galaxies can now
be studied at redshifts as high as $z=1$ (Vogt et al, 1996), while
absorption line studies of QSOs offer indirect evidence that some disc
galaxies were in place at redshifts as high as $z=3.15$ (Lu, Sargent
and Barlow 1997). More likely the Galaxy was formed at a redshift of
between 1 and 2 (Lacey et al 1997).

  Our results weakly indicate that the inner disc is older than the
disc at the solar circle, which naturally turns our attention to the
center of the Galaxy.  There is some evidence that the bulge is older
than the halo (Lee, 1992) from its RR Lyrae stars, although the
results of the MACHO survey indicate that RR Lyrae stars
preferentially tracing the inner halo, rather than bulge (Alcock et al
1997). It may be possible to place age limits on the bulge from the
stars in the OGLE micro-lensing data (Paczynski and Stanek 1997).

\section{Conclusions}

The main conclusions of our work are:

\begin{enumerate}

\item Using the accurate absolute magnitudes of clump giants in the
Hipparcos CMD for which metallicities are available from H{\o}g \&
Flynn (1997), we have obtained an accurate age for the Galactic
disc. We find the Galactic disc to be $11 \pm 2$ Gyr.

\item An age of 11 Gyr for the Galactic disc implies that it formed
rapidly after the most metal rich globular clusters. In conjunction
with the new ages derived for globular clusters by Jimenez et al
(1996), the delay between the formation of the halo and disc was 2-3
Gyr.

\item The reddest stars in the red clump can be used to determine the
metallicity of stellar populations between 2 and 12 Gyr. We have shown
that the metallicity for the Galactic disc inferred using this method
is in good agreement with the one measured by H{\o}g \& Flynn (1997).

\item The morphology of the red clump is well modelled by a Reimers
mass loss law, with $\eta$ changing from 0.0 to 0.8.

\end{enumerate}

\end{document}